\documentclass[]{interact}

\usepackage{epstopdf}

\usepackage[numbers,sort&compress]{natbib}
\bibpunct[, ]{[}{]}{,}{n}{,}{,}
\RequirePackage{amsthm, amsmath, amssymb, mathtools, amssymb, natbib}
\RequirePackage{graphicx, xcolor, makecell}
\RequirePackage{subfigure}

\renewcommand\bibfont{\fontsize{10}{12}\selectfont}

\theoremstyle{plain}

\theoremstyle{definition}

\theoremstyle{remark}

\newcommand{\bx}{\mathbf{x}}
\newcommand{\bX}{\mathbf{X}}
\newcommand{\by}{\mathbf{y}}
\newcommand{\bz}{\mathbf{z}}
\newcommand{\bs}{\mathbf{s}}

\def\nc{\normalcolor}

\newcommand{\ech}{\nc  }          

\setcitestyle{authoryear,open={(},close={)}}

\begin{document}

\articletype{RESEARCH ARTICLE}

\title{Prediction and Model Evaluation for Space-Time Data}

\author{
\name{G.~L. Watson\textsuperscript{a}\thanks{CONTACT: D. Telesca. Email: donatello.telesca@ucla.edu}, C.~E. Reid\textsuperscript{b}, M. Jerrett\textsuperscript{c} and  D. Telesca\textsuperscript{a}}
\affil{\textsuperscript{a} University of California Los Angeles, Department of Biostatistics, \textsuperscript{b} University of Colorado, Boulder, Department of Geography, \textsuperscript{c} University of California, Los Angeles, Department of Environmental Health Sciences.}
}

\maketitle

\begin{abstract}
Evaluation metrics for prediction error, model selection and  model  averaging on space-time data are understudied and poorly understood. The absence of independent replication makes prediction ambiguous as a concept and renders evaluation procedures developed for independent data inappropriate for most space-time prediction problems. Motivated by air pollution data collected during California wildfires in 2008, this manuscript attempts a formalization of the true prediction error associated with spatial interpolation. We investigate a variety of cross-validation (CV) procedures employing both simulations and case studies to provide insight into the nature of the estimand targeted by alternative data partition strategies. Consistent with recent best practice, we find that location-based cross-validation is appropriate for estimating spatial interpolation error as in our analysis of the California wildfire data. Interestingly, commonly held notions of bias-variance trade-off of CV fold size do not  trivially apply  to  dependent data, and we recommend leave-one-location-out (LOLO) CV as the preferred prediction error metric for spatial interpolation. 
\end{abstract}


\section{Introduction}

Space-time interpolation is an important task in air pollution epidemiology for estimating exposure over a continuous space-time domain within which pollutants have been observed at discrete locations and times. This task is often cast as an exercise in prediction, but in certain contexts goes by other names, including downscaling, data fusion, land use regression and infill prediction. A wide variety of deterministic, statistical and machine learning models have been used to interpolate spatial or space-time data, indicative of the ongoing and diverse interest in the problem~\citep{van2015high, reid2015spatiotemporal, watson2019machine, berrocal2020comparison}. The selection, tuning and evaluation of interpolation models typically uses cross-validation~(CV) to estimate prediction error. CV is appealing, because it is easy to perform and allows for the comparison of different classes of prediction models.

While CV is widely used, its behavior in the context of space-time dependence is inadequately understood. Efron famously characterized CV estimation of prediction error for independent data~\citep{efron1983estimating}, but little has been done to extend this formalization to the case of dependent data, for which independent replication in not observable. Space-time dependence complicates model evaluation in two ways: (1) evaluation procedures developed for independent data are generally inappropriate, because they underestimate the true error~\citep{roberts2017cross}; (2) multiple types of prediction are possible due to different relationships between the training data and the data to be predicted, (e.g., forecasting as opposed to spatial interpolation), whereas prediction is unambiguous in the context of independent data. There is growing awareness in the literature that independent model evaluation procedures should not be naively applied to models fit to space-time data, prompting various alternatives. Motivated by air pollution data collected during California wildfires in 2008, we provide a formalization of the true prediction error associated with spatial interpolation and investigate a variety of CV procedures using simulations and case studies.  Particulate and gaseous air pollution from wildfires often display extreme spatiotemporal heterogeneity, which makes this a particularly germane example with which to interrogate prediction error.

We restrict our discussion to approaches that estimate prediction error per~se, notably excluding information criteria, which are popular tools for model comparison. We note that some information criteria require a likelihood to be computed, which may not exist for some prediction algorithms, and that the large number of proposed information criteria makes a thorough discussion infeasible. 
We also operate without concern for consistency of model selection, focusing on evaluation metrics that estimate prediction error. In a model selection context our goal is to select the best predicting model, i.e., that with the lowest expected prediction error. We do not require that such a procedure converge to the true model if indeed a true model were to exist.

The remainder of the paper proceeds as follows. Section \ref{sec:data} introduces a motivating case study on wildfire pollution; Section~\ref{sec:predError} reviews prediction error estimands and estimators first for independent data and then for space-time dependent data, formalizing the true prediction error associated with spatial interpolation; Sections~\ref{SimulationSection} and~\ref{CaseStudySection} investigate the behavior of these estimators using a battery of simulations and two case studies; and Section~\ref{sec:discussion} summarizes estimator performance and offers recommendations for best practices.

\section{Motivating Application} 
\label{sec:data}

Daily measurements of ozone (8-hour maximum) and  24-hour average  PM$_{2.5}$ (particulate matter smaller than 2.5$\mu$m in diameter) were recorded at monitors across the state of California from May 1, 2008, through September 30, 2008. Within this time period a large complex of wildfires burned for several weeks in northern California, shrouding much of the state in smoke and exposing its populace to toxic pollutants.  

Inferring the health consequences of exposure to these pollutants is typically done in two stages: (1) a space-time prediction model trained on the monitor data and fine-resolution covariates  translates pollution measurements from the coarse space-time resolution of the monitors to the much finer spatial resolution of health outcome data, and (2) health outcomes are regressed on predicted pollution exposure adjusting for potentially confounding covariates. The two-stage approach breaks a challenging problem into two more manageable tasks, although uncertainty or inaccuracy in exposure predictions can induce bias in the subsequent epidemiological analysis~\citep{szpiro2013measurement}. Avoiding or correcting for this bias is an active area of research.  This manuscript concerns the first stage.

\section{Prediction Error: Estimands and Estimators} 
\label{sec:predError}

\subsection{The Case of Independent Replication}
\label{sec:review}

Let $\bz  =  \{z_1,\ldots, z_n\}$  denote  $n$ independent and identically distributed (iid) samples from an unknown probability distribution $F$, such that $z_1,\ldots, z_n \sim_{iid} F$. We assume $z_i = \{y_i, \mathbf{x}_i\}$, where $y_i \in \mathbb{R}$ denotes a real-valued outcome, and $\mathbf{x}_i \in \mathbb{R}^p$ denotes a random vector of $p$ covariates.   In this context, a common framework often used to formalize the notion of prediction error relies on the existence of an independent and identically distributed test point $z_0 = \{y_0, \bx_0\} \sim F$. 

Let $\eta(\bx_0, \bz): \mathbb{R}^{np + n + p} \mapsto \mathbb{R}$ define a prediction rule that maps 
the training data $\bz$ and a covariate vector $\bx_0$ to a real number that serves as a prediction for $y_0$. Discrepancies between 
$y_0$ and its predicted value may be quantified by a 
loss function, $Q\{y_0, \eta(\bx_0, \bz)\}: \mathbb{R}^2 \mapsto \mathbb{R}^{+}$, mapping $y_0$ and its prediction $\eta(\bx_0, \bz)$ to a non-negative real number.  This loss function is easily generalized to quantify prediction across multiple test points, $\bz_0 = \{z_{01}, ..., z_{0m}\}$, by summing the loss at individual points, i.e., $Q\{\by_0, \eta(\bX_0, \bz)\} = \sum_{j=1}^m Q\{y_{0j}, \eta(\bx_{0j}, \bz)\}$. 
Given 
 the training data $\bz$, the \emph{true error} of a prediction rule trained on $\mathbf{z}$ is 
\begin{equation} \label{IndepTrueErr}
  Err \coloneqq E_{F} [ Q\{y_0, \eta(\bx_0, \bz)\}];
\end{equation}
the expected loss incurred by $\eta(\cdot)$, where 
 $E_{F}$ is an expectation taken with respect to probability distribution $F$, over the state space of $z_0$ \citep{efron1983estimating}.


We operate under the assumption that $F$ is unknown, as is typical in many applications. In this context, the expected loss in equation \ref{IndepTrueErr} cannot be computed in closed form, and the definition of estimators of $Err$ with good operating characteristics becomes a surprisingly formidable problem \citep{Efron:2004}. Some progress is often made by constructing estimators which rely on the separation of training and testing through data partitions. 

For example, $K$-fold cross-validation (CV) \citep{stone1974cross} 
partitions the training data $\mathbf{z}$ into $K$ disjoint subsets often referred to as folds, $\mathbf{z}^{(1)} = \{\mathbf{y}^{(1)}, \mathbf{x}^{(1)}\},\ldots,\mathbf{z}^{(K)} = \{\mathbf{y}^{(K)}, \mathbf{x}^{(K)}\}$, usually of equal or nearly equal size, where $K$ is an integer between 2 and $n$, so that $\mathbf{z} = \cup_{k=1}^K \mathbf{z}^{(k)}$, and $\mathbf{z}^{(k)} \cap \mathbf{z}^{(\ell)} = \varnothing$ for $k \neq \ell$. The prediction rule $\eta$ is trained $K$ times, each time excluding one subset, with the withheld subset serving as a test set on which to evaluate the prediction rule trained on the rest of the data.  The $K$-fold CV estimator of prediction error averages the loss function evaluated over each of the withheld subsets, 
\begin{equation}  \label{KfoldCV} 
  \widehat{err}^{(CV_K)} \coloneqq \frac{1}{K} \sum_{k = 1}^K  \frac{1}{n_k} Q\left\{\mathbf{y}^{(k)}, \eta\left(\mathbf{x}^{(k)}, \mathbf{z}^{(-k)}\right)\right\},
\end{equation}
where $\mathbf{z}^{(-k)}$ is the training data excluding the $k$-th fold, i.e., $\mathbf{z}^{(-k)} \coloneqq \mathbf{z} \setminus \mathbf{z}^{(k)}$, and $n_k$ is the number of observations in $\mathbf{z}^{(k)}$. 


In a similar fashion, one may rely on the bootstrap \citep{efron1979bootstrap}. The procedure generates $B$ data sets, $\mathbf{z}^*_1, \ldots, \mathbf{z}^*_B$, by resampling $n$ observations with replacement 
from the training data $\mathbf{z}$. For each bootstrap dataset $\bz_b^*$ $(b=1,\ldots,B)$, let $\bz_{-b}^* \coloneqq \bz \setminus \bz_b^*$  denote the training data withheld from the $b$-th bootstrap sample. 
The \emph{out-of-bag} estimator of $Err$ is defined as follows, 
\begin{equation*}
  \widehat{err}^{(boot)} \coloneqq \frac{1}{B} \sum_{b=1}^B Q\left\{\mathbf{y}_{-b}^*, \eta\left(\mathbf{x}_{-b}^*, \mathbf{z}_b^*\right)\right\}.
\end{equation*}
When the training data, $\mathbf{z}$, are independent samples from $F$ and have finite, positive variance, the bootstrap estimator converges in distribution to the true error \citep{bickel1981some}.  Further refinements of these two basic estimators are reported in \cite{zhang1995assessing} and  \cite{efron1997improvements}.

A more direct estimator of $Err$ is often used in practice when $n$ is large. In this setting, one relies on a validation set $\mathbf{z}^*$ consisting of $m$ samples from $F$ independent of $\mathbf{z}$. The expectation in equation \ref{IndepTrueErr} may be approximated by averaging the loss function over the validation set, yielding the \emph{validation error}, 
\begin{equation*}
  {err}^* \coloneqq \frac{1}{m} \sum_{i=1}^m Q\left\{y_i^*, \eta\left(\mathbf{x}_i^*, \mathbf{z}\right)\right\},
\end{equation*}
where $\{ y_i^*, \mathbf{x}_i^*\}$ is the $i$-th validation sample. Because the validation set consists of independent samples from $F$, it is an unbiased estimator of the true error with a variance dependent on the number of validation samples $m$.

Reliance on the basic idea of training a prediction rule on a subset of the data and evaluating it on a different (i.e., disjoint or independent) subset might be proscribed in the absence of independent replication. The use of cross-validation is, however, prevalent in several applications of machine learning to spatial and space-time data. Therefore, in the following section we attempt to clarify the role of data-partition strategies in space-time analysis.

\subsection{Prediction Error Estimands for Space-Time Data}
\label{sec:ErrorDep}

We characterize the sampling structure of space-time data using the formalism of marked point processes~\citep{moller2007modern}.  Specifically, given a bounded spatial region $\mathcal{S}\subset\mathbb{R}^2$, we assume that data is observed at a spatial point pattern $\bs = \{s_1, s_2,\ldots, s_n\} $, $s_i \in \mathcal{S}$,  $(i=1,2,\ldots,n)$. Associated with each spatial location $s_i$, we observe a time series of real-valued outcomes $\by(s_i) = \{y(s_i, t_1), \ldots, y(s_i, t_{m_i})\}$, as well as a matrix of real-valued covariates $\bX(s_i) = \{\bx(s_i,t_1), \ldots, \bx(s_i,t_{m_i})\} \in \mathbb{R}^{p\times m_i}$, where $m_i$ is the length of the time series observed at $s_i$. The collection of outcomes and covariate information at location $s_i$ is denoted as $z(s_i)\coloneqq \{\by(s_i), \bX(s_i)\}$ and may be formally characterized as a point-process mark at location $s_i$.

Within this framework, a space-time dataset is therefore composed of a spatial point pattern $\bs$ with associated marks $\bz(\bs) = \{z(s_1),\ldots,z(s_n)\}$, and sampling distribution $F\{\bs, \bz(\bs)\} \coloneqq F_s(\bs)F_{z\mid s}\{\bz(\bs)\mid\bs\}$,  which may be represented hierarchically as the marginal distribution of the point pattern $F_\bs$ times the conditional distribution of the marks $F_{\bz\mid \bs}$.

In this setting, the notion of prediction is inherently ambiguous until one qualifies how a test data point is to be obtained.  In particular,  \emph{spatial interpolation} is formally interpreted as the conditional prediction at location $s_0\sim F_{s}$ of $\by(s_0)$, given $\bX(s_0)$ and $\bz(\bs)$.   Let $\eta\left(\bX(s_0), \bz(\bs)\right)$ represent a prediction rule, and $Q\{\by(s_0), \eta\left(\bX(s_0),\bz(\bs)\right)\}$  be a loss function, both 
analogous to our definitions in Section \ref{sec:review}. Given a training marked spatial pattern $\bz(\bs)$, the \emph{interpolation error}  of $\eta(\cdot)$ trained on $\bz(\bs)$, is naturally defined as  
\begin{equation} \label{InterpError}
    Err_s \coloneqq E_{F\{s_0, z(s_0)\}} \left[Q\{\by(s_0), \eta(\bX(s_0),\bz(\bs))\}\right];
\end{equation} 
the expected loss incurred by $\eta(\cdot)$, where $E_{F\{s_0, z(s_0)\}}$ is an expectation taken with
respect to probability distribution $F$, over the joint state space of $\{s_0, z(s_0)\}$.

We note that, in the definition of $Err_s$, the spatial point pattern distribution $F_s(\cdot)$, need not coincide with the one defining the observed point pattern $\bs$. It could be useful, for instance, to use a sampling schedule for $s_0$ proportional to the density of the population at risk. In this example, the use of exogenous information may be critical when the goal is one of providing spatial interpolation for environmental stressors, because predictive accuracy is much more important in populated areas.

\subsection{Prediction Error Estimators for Space-Time Data}

The formalism introduced in Section  \ref{sec:ErrorDep} will serve as a guiding framework for the characterization of structured data partition strategies defining prediction error estimators for space-time data. 

Standard $K$-fold~CV (equation~\ref{KfoldCV}) is still commonly used on space-time data~\citep{reid2015spatiotemporal, guo2017estimating}  despite acknowledgement in the spatial statistics community that its estimates are too optimistic~\citep{watson2019machine, roberts2017cross}.  In light of the estimand defined in (\ref{InterpError}), unstructured data partitions are unlikely to exclude data associated with specific locations, making it very likely that locations represented in the test set also appear in the training data. 
Rather than targeting the spatial interpolation error, na\"{i}ve CV estimators appear to target the \emph{imputation error}, i.e. the expected loss incurred by a prediction rule $\eta(\cdot)$ for unobserved time points at observed locations. 

We are not the first to point out these problems with standard  $K$-fold~CV. For example, \cite{roberts2017cross} 
advocate for the use of spatial or space-time blocking as structured data partition strategies targeting prediction errors in spatial and space-time data. Specifically, the spatial (or space-time) domain is divided into rectangular (or cubic) regions that serve as a data-partition unit. The sizing of data blocks  aims to  reduce or eliminate the impact of spatial (or space-time) dependence on estimates of prediction error, in an effort to replicate estimators defined in iid settings. This approach thus appears to target a form of \emph{replication error}, i.e. the expected loss incurred by a prediction rule $\eta(\cdot)$ for a new realization of the marked space-time process, which is critically different from the spatial interpolation error defined in~(\ref{InterpError}). 

Alternatively, spatially buffered CV has been proposed for mitigating the effect of spatial dependence for prediction error estimation and model selection~\citep{le2014spatial, pohjankukka2017estimating}. These strategies pick one or more spatial locations for each test fold and train the model on observations located beyond a buffer distance from the test locations. The usual goal is to pick a buffer distance that excludes observations from the training data that have residual spatial correlation with the test data.  This is the same goal as spatial blocking and appears to similarly target the replication error. Smaller buffers that do not entirely remove spatially dependent training data appear to target a \emph{spatial extrapolation error}, i.e., the error associated with predicting at a spatially distant (but not independent) location.

The key to defining a structured CV strategy for the estimation of the interpolation error in (\ref{InterpError}), is naturally implied by the sampling structure of the marked point process introduced in Section \ref{sec:ErrorDep}.  More precisely, let $\bz\left(\bs_{(i)}\right)$ be the set of observations across all locations, except for $s_i$. We define the leave-one-location-out  cross-validation (LOLO-CV) estimator of $Err_s$ as, 
\begin{equation} \label{LOLOestimator}
\widehat{err} _s^ {(LOLO)} \coloneqq  \frac{1}{n} \sum_{i = 1} ^ {n} Q\left\{\by(s_i), \eta\left(\bX(s_i), \bz\left(\bs_{(i)}\right)\right)\right\},
\end{equation}
in which the CV fold left out of model training is the entire time series of observations taken at a particular location. This is analogous to the leave-one-out CV estimator in the iid setting, except the CV resampling unit is no longer individual observations but rather every observation taken at an individual location.
Defining each CV fold as the observations taken at one location ensures that predictions are always being evaluated at a location excluded from the data on which the prediction rule $\eta(\cdot)$ was trained. This intuitively mimics interpolation in which the goal is to predict the time series at an unobserved location. This intuition is presumably the motivation for a number of authors using LOLO-CV \citep{keller2017measurement, watson2019machine}, even though, to our knowledge, our analysis is the first to attempt a formalization of the target estimand in (\ref{InterpError}).

LOLO-CV can be generalized to partition data over multiple locations.  Specifically, we partition the observed point pattern $\bs$ into 
$K$ disjoint location-subsets or spatial folds, $\bs^{(1)},\ldots, \bs^{(K)}$, with associated marks $\bz(\bs^{(1)}), \ldots, \bz(\bs^{(K)})$, so that $\bs = \cup_{k=1}^K \bs^{(k)}$, and $\bs^{(k)} \cap \bs^{(\ell)} = \emptyset$, for $k\neq \ell$. The $K$-fold leave-location-out ($LLO_K$)-CV estimator of $Err_s$ is defined as, 
\begin{equation*}
     \widehat{err}_s ^ {(LLO_K)} \coloneqq   \frac{1}{K}\sum_{k=1}^K\frac{1}{n_k}\sum_{i\in                   \bs^{(k)}} Q\{\by(s_i), \eta(\bX(s_i), \bz(\bs^{(-i)}))\},
     \end{equation*}
where $\bz(\bs^{(-i)})$ is the set of observations at all locations excluding the spatial points in $\bs^{(i)}$, and $n_k$ is the number of locations in spatial fold k.  

 Setting $K=n$ yields the leave-one-location-out (LOLO) cross-validation estimate of prediction error, $\widehat{err}_s ^ {(LOLO)}$ in (\ref{LOLOestimator}). LLO$_{K}$-CV trains and evaluates $K$ models so that when $K$ is smaller than $n$, LLO$_{K}$ is computationally expedient compared to LOLO-CV.  Setting $K=10$ is a common choice, presumably on analogy to the iid setting \citep{lee2016spatiotemporal}.

As in the independent case, there is a bias-variance tradeoff between these estimators. In the context of independent data, the LOO estimator has lower bias, but higher variance than  $K$-fold~CV estimators for $K > 1$. In the dependent setting, however, removing locations from the data can result in hard to predict sources of bias, and the standard wisdom that suggests 10-fold~CV as a rule of thumb estimator~\citep{kohavi1995study} for balancing bias and variance may not apply. Estimator bias depends on the dependence structure, the density of observed locations and the number of observations in each time series. We elucidate some of these aspects in our simulation studies, showing that LOLO~CV seems to be the best for typical studies in which the observed data include  a relatively small number of locations and a relatively large number of time points.

LOLO-CV and LLO$_K$-CV use the empirical distribution of locations $\hat{F}(\mathbf{s})$ in computing the average loss of $\eta$. The observed locations provide an estimate of the spatial intensity function, which weights the average loss more heavily in areas with more observed locations. If prediction error is to be taken as the expectation over a uniform spatial field, then using the empirical spatial intensity may introduce sampling bias \citep{diggle2010geostatistical, gelfand2012effect}. In epidemiological applications, however, it is often desirable to define prediction error over a non-uniform sampling intensity, with the preferential sampling pattern of pollution monitors providing critical information into the relative importance of different locations. If it is desirable to define prediction error as the expected loss over a different location sampling intensity, it is foreseeable to apply  corrections to $\hat{err} ^ {(LOLO)}$ and $\hat{err}^{(LLO_K)}$. This could be used to estimate the expected loss uniformly across the spatial region or to incorporate ancillary information, for example, direct estimates of population density if prediction error were to be more heavily weighted where the population were more dense. These corrected estimators are not investigated in this manuscript.


\section{Simulation Study} \label{SimulationSection}

To illustrate the behavior of these prediction error estimators in the context of dependent data, we perform a series of simulation studies. Each simulation is conducted over a dense space-time grid with a small subset of the grid locations sampled as observed locations. Six covariates and the outcome were simulated over the space-time grid. The covariates were simulated from a Gaussian process with a separable covariance function and a sinusoidal mean function. The outcome was simulated from a Gaussian process with a separable covariance function and a mean function that depended on the first three covariates with linear, sigmoidal and threshold effects respectively. The remaining three covariates are spurious. The precise form of each mean and covariance function is detailed  in section~S2  of the supplemental material.

For a given model, the true prediction error is computed as the mean square error (MSE) of predictions made over the unobserved portion of the grid by a model trained on the observed data. Each estimator of prediction error is then computed over the observed data and compared to the true prediction error. To help elucidate the impact of sample size and dependence structure on estimator behavior and motivated by air pollution data sets, we chose two different numbers of observed monitors, 50 and 150, and two sets of spatial dependence parameters, corresponding to low and moderate dependence. Table~S3 in the supplement 
lists the parameters of each simulation scenario. Points near the boundary of the grid have fewer nearby points, which can result in unexpected edge effects. To avoid this, only points sufficiently far from the grid boundary were eligible for selection as observed locations. We evaluated four estimators of prediction error: (1) naive 10-fold~CV, (2) LLO$_{10}$~CV, (3) LOLO~CV and (4) buffered leave-location-out CV
on three different models: linear regression, random forest and space-time Kriging. These models were selected as examples of the three main classes of models used in the literature: (1) simple regression models, (2) machine learning models with flexible mean structures that assume independent data, and (3) Kriging models with rigid mean structures but relatively flexible covariance models. These three models are far from a comprehensive list and are intended to illustrate the behavior of different CV procedures on simple, but different classes of prediction strategies. 

\subsection{Simulation Results \& Discussion} 

\begin{figure}	
		\centering
		\begin{tabular}{c}
                   \includegraphics[width=4in]{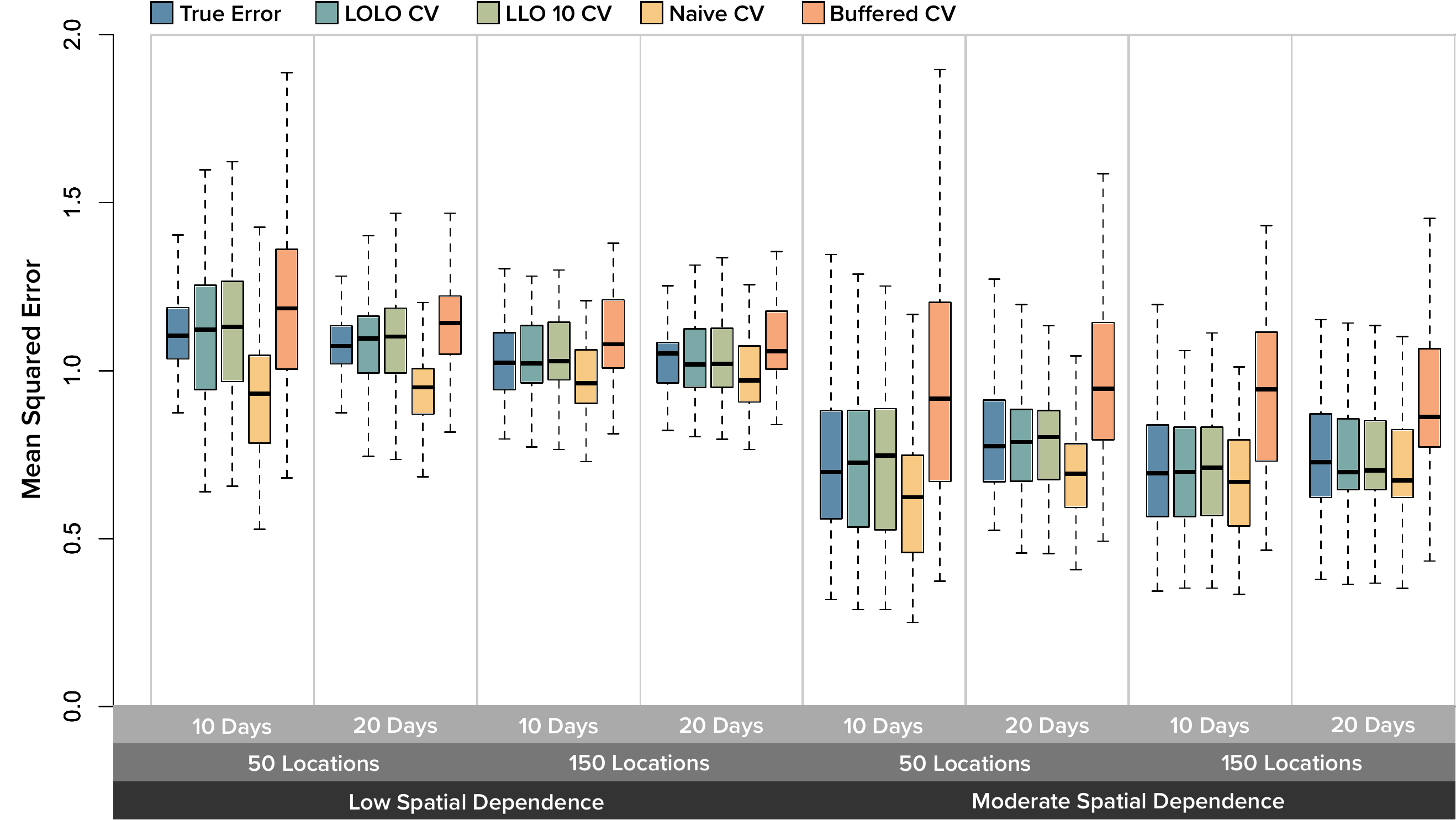} \\
		  (a) Simple Linear Regression \\
		  \includegraphics[width=4in]{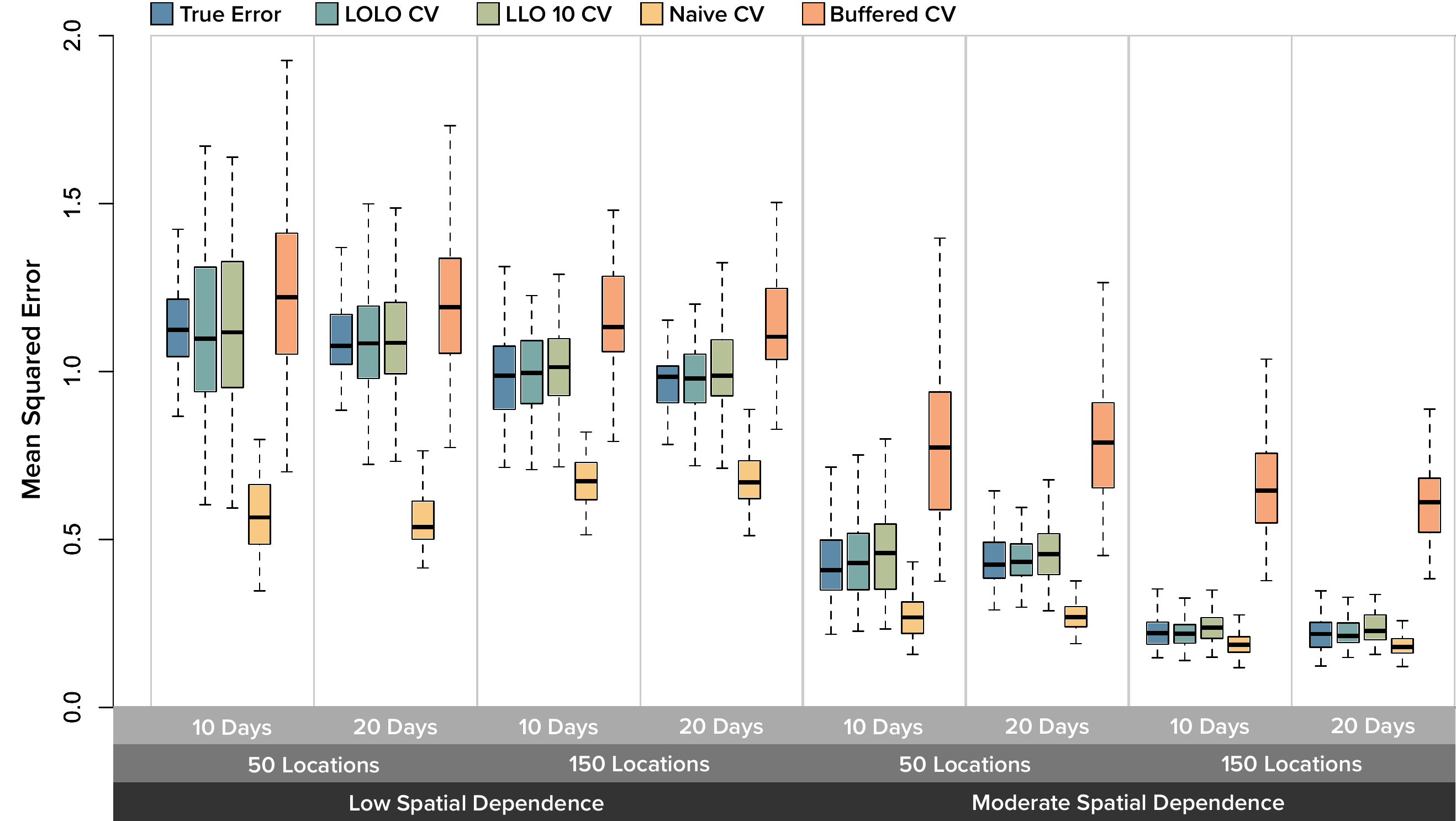} \\
		  (b) Random Forest \\
                     \includegraphics[width=4in]{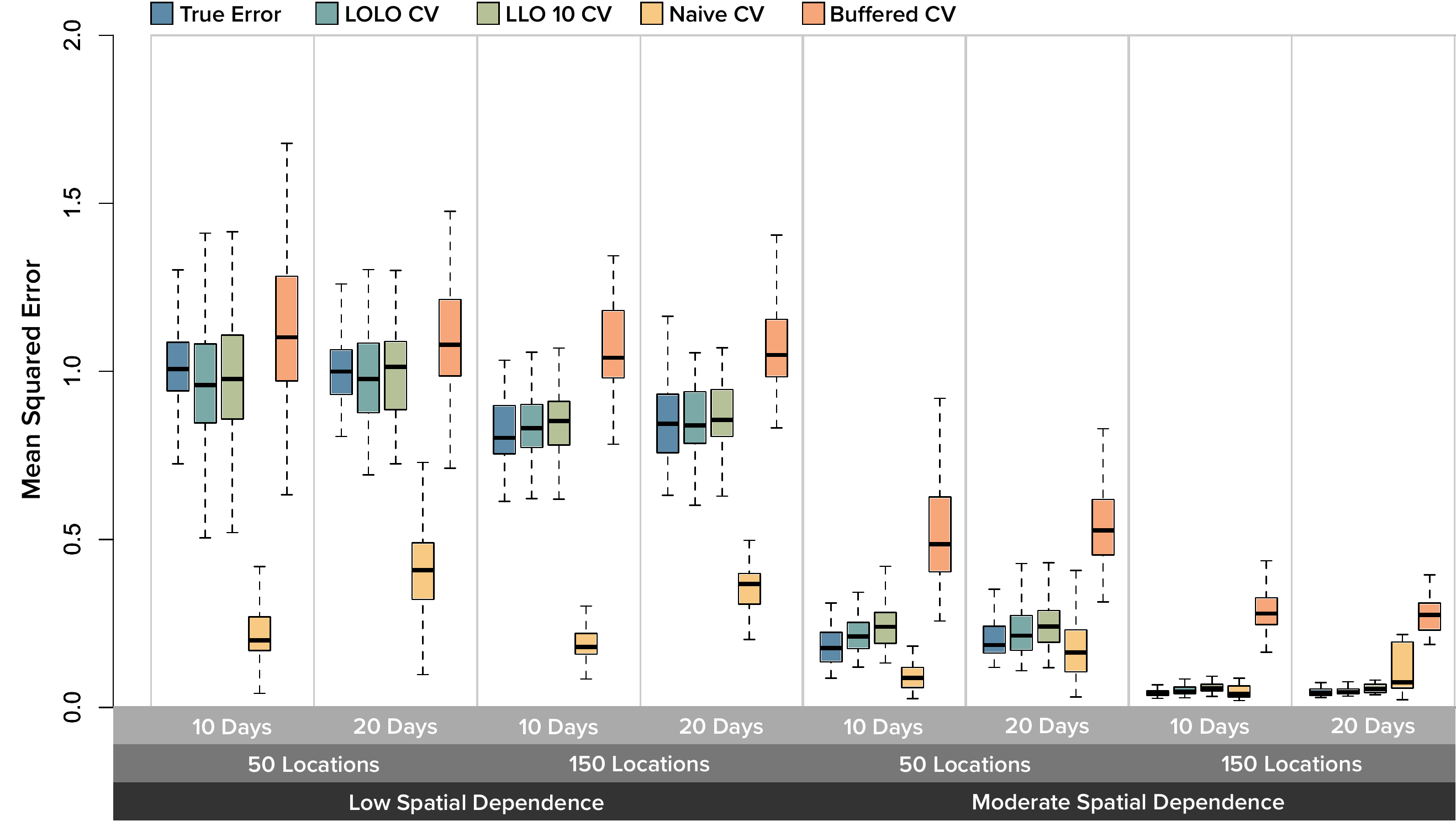} \\
                   (c) Kriging 
                 \end{tabular}
		\caption{Boxplots of the true error and its estimators for each model across the 100 replicates of each simulation scenario.} 		
		\label{fig:SimResults}
\end{figure}

Figure~\ref{fig:SimResults} depicts the true error and four estimates for each model across the 100 replicates of each simulation scenario.  Across all three models, LOLO~CV (turquoise) and LLO$_{10}$~CV (green) appear to target the true error (blue). 

Naive 10-fold CV (yellow) is very optimistic for random forest and Kriging, especially when spatial dependence is low or relatively few locations are observed. The information provided by observations at other time points at the location of the data point being predicted is apparently very informative for models flexible enough to exploit it.  Naive CV is less optimistic for simple linear regression, because it is much less flexible than random forest or Kriging and consequently worse at exploiting this information. 

The optimism of naive 10-fold CV is less pronounced as the number of observed locations or the spatial dependence increases for all 3 models. In these cases, the other observed locations are more predictive of the test location, making naive CV's access to observations for other timepoints at the test location less advantageous and thereby reducing its optimistic bias.  Interestingly, the naive 10-fold CV estimates of prediction error for Kriging become less stable for longer time series despite the corresponding additional data, suggesting that naive 10-fold CV is particularly unreliable for techniques that rely upon the dependence structure for prediction. 

The LOLO estimates of prediction error tend to be slightly smaller and less biased (i.e., closer to the true error) than LLO$_{10}$ for two reasons. First, LLO$_{10}$  generally uses smaller data sets for training because its folds are larger, and the additional training data available to LOLO reduces its bias. This result is well-known in cross-validated studies of independent data. Second, withholding more locations further hinders the LLO$_{10}$ training models because it removes the information they provide on other locations due to the spatial dependence between locations. This does not occur in the independent setting where there is no dependence between observations. 

This effect makes the bias-variance trade-off of CV fold size more complicated than in the independent case. For independent data, as the fraction of the data used in the training data increases, the bias decreases, but the variance increases. Leave-one-out CV includes all but 1 observation in the training data and has smaller bias but greater variance than 10-fold CV, which includes 90\% of the data in the training data. For data exhibiting space-time dependence, this relationship need not hold. Whether LOLO CV has higher variance than LLO$_{10}$  CV depends on the dependence structure, the number of observed locations and the length of the time series observed at each location. 

Spatially buffered CV (orange) is pessimistic (i.e., biased upward) in all cases, which is consistent with the suggestion that it may target a spatial extrapolation error rather than the interpolation error. For a fixed buffer size, its pessimism is greater for higher spatial dependence, since the locations within the buffer that are withheld from the training model are more highly correlated with the observations being predicted. Its pessimism is lower when more locations are observed, because the additional locations beyond the buffer may partially make up for the loss of information associated with withholding observations within the buffer.  

\begin{figure}	
		\includegraphics[width=5in]{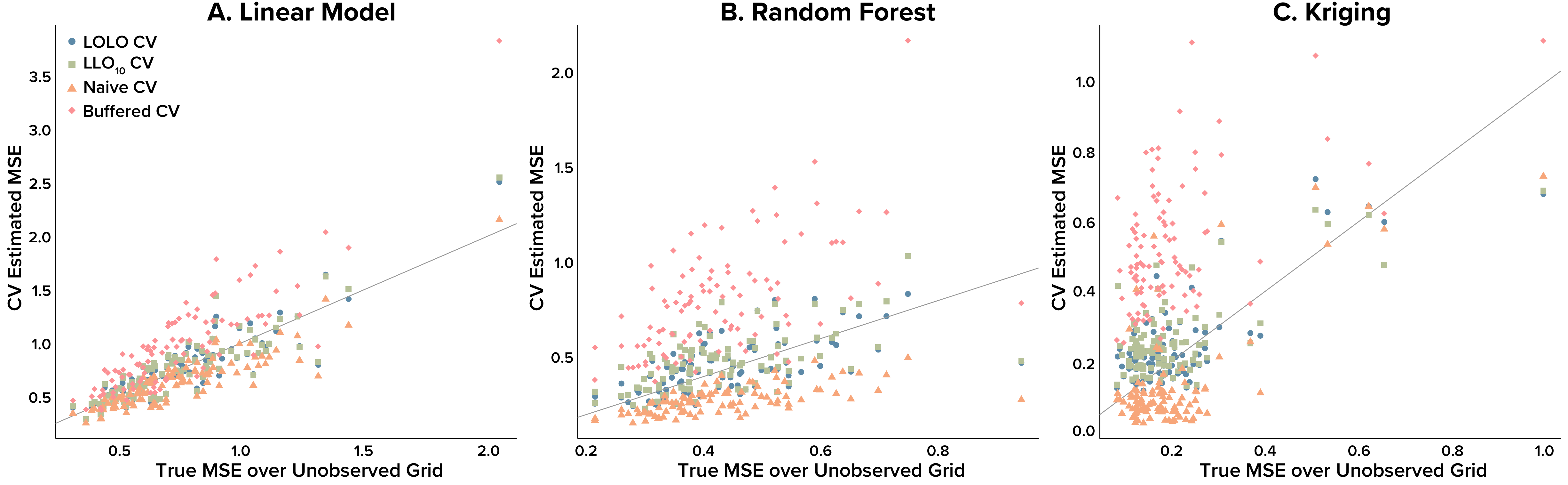}
		\caption{ Scatter plots of the true MSE computed over the unobserved grid against CV estimates of MSE for (a) Simple Linear Regression, (b) Random Forest, and (c) Kriging.  The grey line depicts $y=x$, i.e., perfect correlation. } 
			\label{fig:SimScatter}
\end{figure}

Figure~\ref{fig:SimScatter} plots MSE against CV estimates of MSE for each of the 3 models for an example simulation scenario.  The CV estimates generally correlate fairly well with the true MSE, suggesting they are not simply accurate on average, but provide useful estimates for a particular data set. The optimistic bias of naive 10-fold CV and the pessimistic bias of buffered CV are evident from their deviation from the line of perfect correlation. As in figure~\ref{fig:SimResults}, these biases are more pronounced for random forest and Kriging on account of their greater flexibility.

\begin{figure}	
		\centering
		\begin{tabular}{cc}
(a)&\hspace{-0.8cm}\includegraphics[width=4.9in]{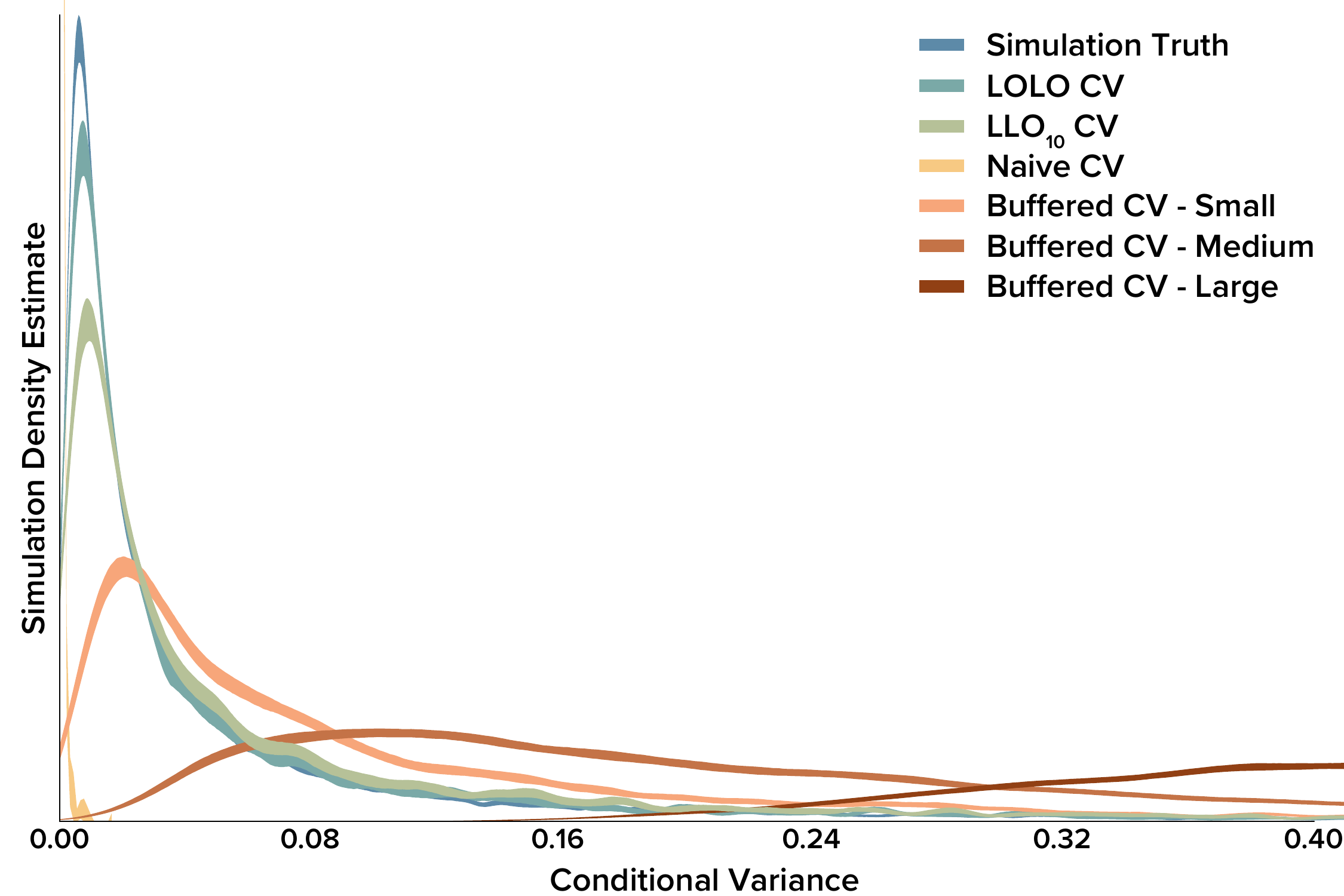} \\[.5in]
(b)&\hspace{-0.8cm}\includegraphics[width=4.9in]{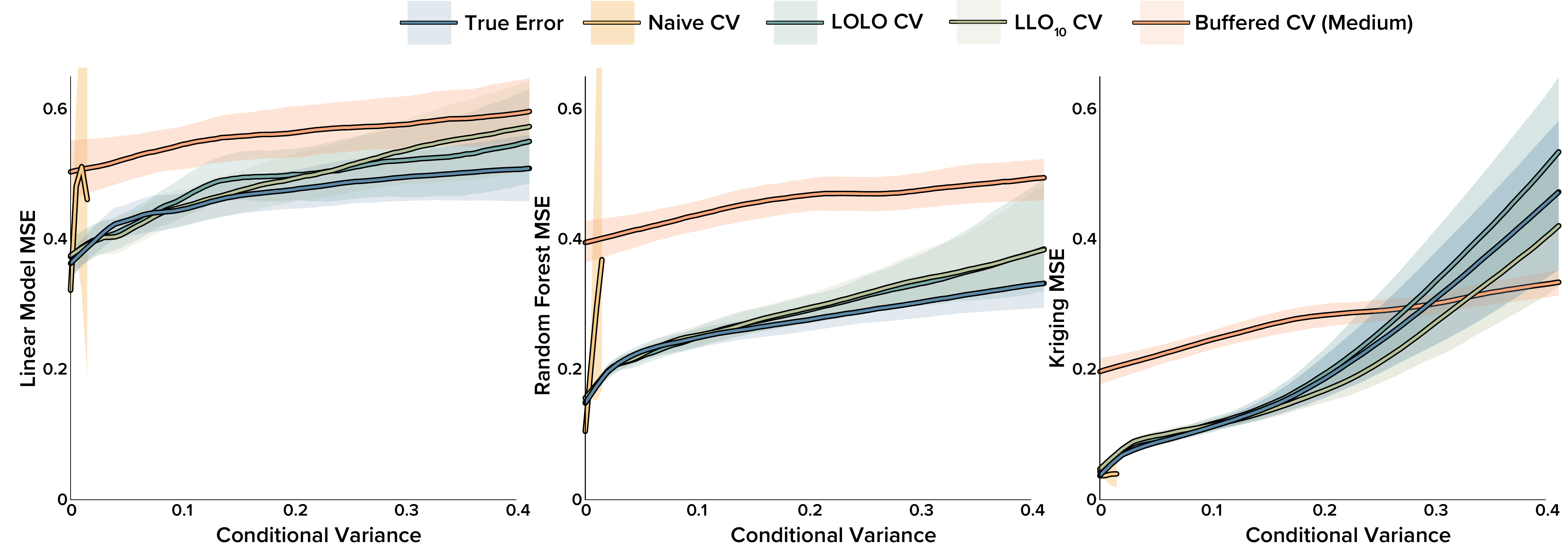} 
		\end{tabular}
	
		\caption{Conditional variance of test points vs. MSE. (a) Density estimates and 95\% pointwise confidence bands 
		of test points given the training data for the 100 replicates of simulation scenario~3. 
		(b)  Median and 95\% pointwise confidence bands for 
		the true and estimated MSE as a function of the conditional variance of test points given the training data for the 100 replicates of simulation scenario~3 (LOWESS smoothed). 
		\ech} 		
		\label{fig:SimConditionalVar}
\end{figure}

It is useful to characterize prediction error as a function of the conditional variance of the outcome to be predicted given the observed data. This conditional variance effectively quantifies the information that the training data provides on the point to be predicted and can be computed exactly in a simulation setting (see section S4 in the supplement for detail). Figure~\ref{fig:SimConditionalVar}
summarizes our findings for simulation scenario~3. In panel (a), we report 
density estimates and 95\% pointwise intervals of the conditional variance for the true error and each cross-validated estimator, using 300 bootstraps of the simulation replicates.

How closely the conditional variance distribution for an estimator matches the true distribution (blue) corresponds to how accurately it estimates the true prediction error. LOLO~CV (turquoise) is most similar to the true error with a slightly greater distribution due to the slightly smaller size of the LOLO training data. LLO$_{10}$~CV (green) is also similar to the true distribution, but the larger fold size increases the conditional variance because of the correspondingly smaller size of the training data. The conditional variance for naive CV is much lower than that of the true distribution and it peaks so sharply near 0 that it is difficult to make out in the figure. This difference in conditional variance describes the optimistic bias of naive~CV: the training data is much more informative for prediction points than the truth, corresponding to a much lower conditional variance and overly optimistic estimates of interpolation error. 

Figure~\ref{fig:SimConditionalVar} includes the conditional variance for three spatially buffered~CV estimators of increasing buffer size. As the buffer size increases so does the conditional variance, because a larger buffer corresponds to more spatially distant and therefore less informative training data. This explains the pessimistic bias of buffered~CV in estimating interpolation error.  
As the buffer size is reduced and less distant locations are included in the training data, buffered~CV approaches LOLO~CV, becoming equivalent when the buffer size is smaller than the closest spatial distance between observations.

	

Figure~\ref{fig:SimConditionalVar} (b) depicts  the median and 95\% pointwise confidence bands for LOWESS (locally weighted smoothing) curves estimating the mean true and estimated errors as a function of conditional variance for each model  over 300 bootstraps of the simulation replicates. As expected, the true error increases with conditional variance. LOLO and LLO$_{10}$~CV also exhibit this relationship, approximating the true error curve quite well for all three models. At larger values of conditional variance there are relatively few data points and consequently greater uncertainty around the estimated curves. This figure shows quite clearly why naive 10-fold CV is such a poor estimator. Because naive 10-fold CV includes observations taken at the same locations as the outcomes being predicted, the conditional variances are very small compared to the conditional variance of the outcome at unobserved locations. 

The medium buffer size spatially buffered CV is also depicted, which consistently overestimates the true error except for Kriging at high conditional variance. This may simply result from the greater uncertainty due to the relatively few observations with high conditional variance, but it may also reflect the sensitivity of Kriging to nearby observations, which are withheld from model training in spatially buffered CV.

Our results in Figure~\ref{fig:SimConditionalVar} (b) also allow for an interesting comparison of predictive models. Kriging predicts very well (i.e., has low MSE) when the conditional variance is small, because it explicitly uses the space-time dependence between data points. Its accuracy falls off substantially, however, as the conditional variance increases. Random forest and simple linear regression ignore this dependence, which reduces their accuracy when the conditional variance is small (especially for simple linear regression), but their performance falls off less quickly at higher conditional variance than does Kriging. 

\section{Case Study} 
\label{CaseStudySection}

Over the weekend of June 20--21, 2008, thousands of lightning strikes in northern California ignited a complex of wildfires that would burn for several weeks, shrouding much of the state in smoke and exposing its populace to toxic pollutants. We set out to showcase how different model evaluation strategies compare, when the goal is producing fine resolution interpolation of pollutant exposure across the state of California. 

More precisely, 186 ozone monitors recorded a total of 25,236 measurements of 8-hour maximum ozone, and 136  PM$_{2.5}$ monitors made 14,247 PM$_{2.5}$ measurements from May 1, 2008, through September 30, 2008, with most monitors recording one observation per day. Linear regression, random forest and universal Kriging (i.e., Kriging with covariates) interpolating prediction models for ozone and PM$_{2.5}$ were fit separately to each week of the data using land use, meteorological, satellite and numerical simulation covariates, which are detailed in table S1 of  the supplemental material. The LOLO, LLO$_{10}$ and naive 10-fold CV estimates of prediction error were computed for each model. To assess the impact of time series length on the prediction error estimators, this procedure was repeated on progressively longer, non-overlapping intervals of 1, 2, 4 and 8 weeks. 

\begin{figure}	
  \centering
    \begin{tabular}{c}
      \includegraphics[width=5in]{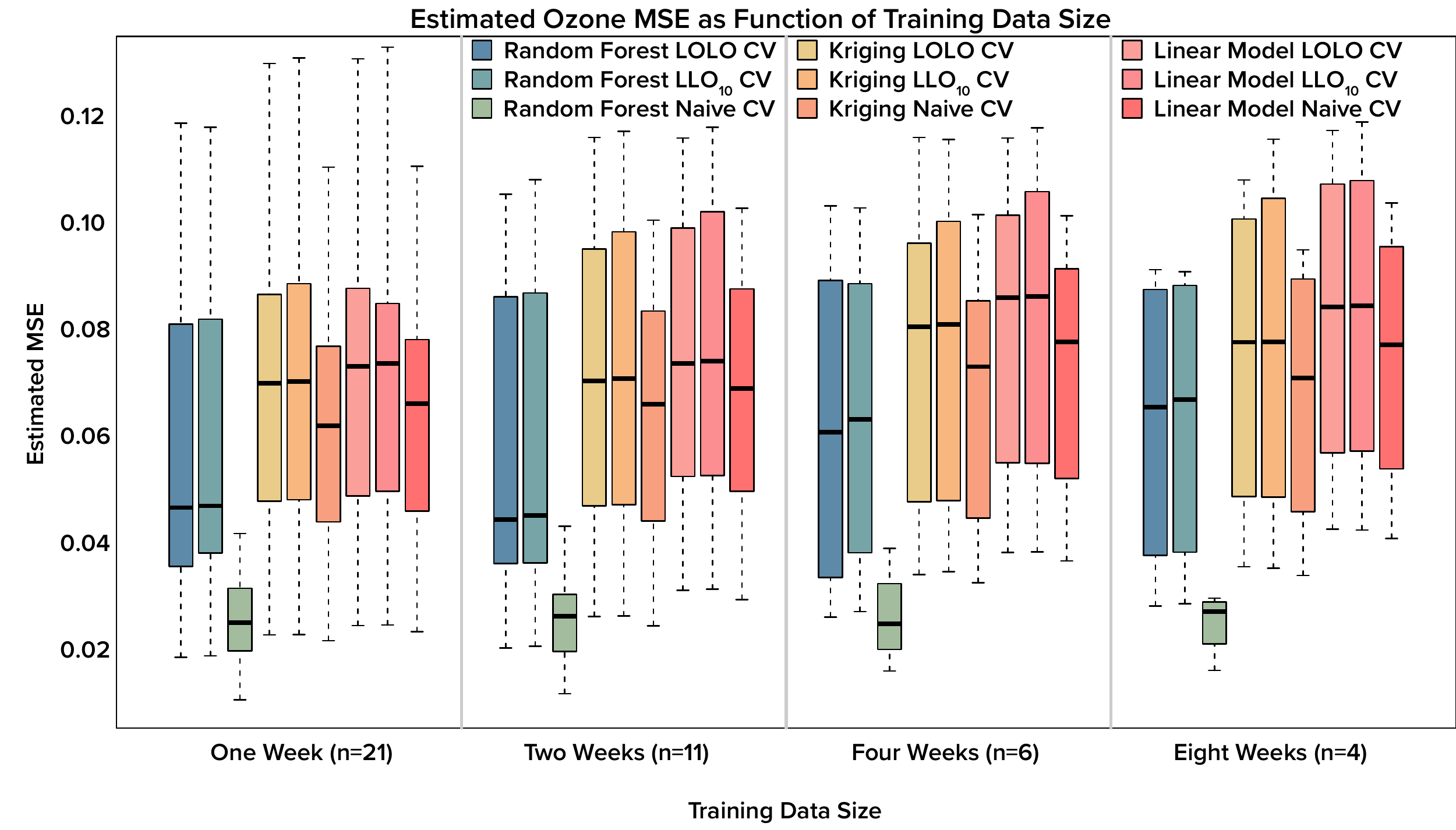}\\
	  (a) Ozone  \\
      \includegraphics[width=5in]{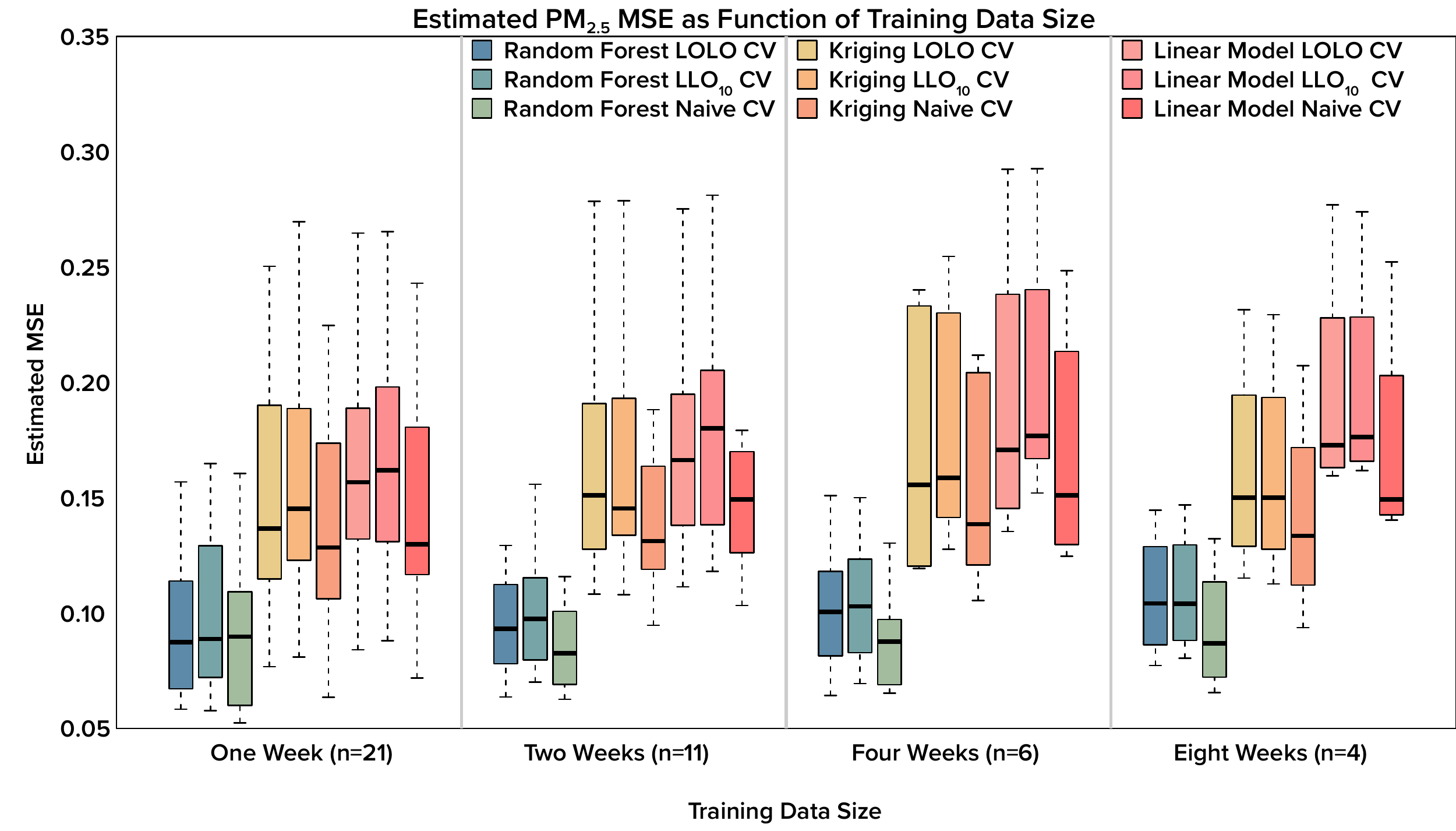}\\
	  (b) PM$_{2.5}$
    \end{tabular}

    \caption{Prediction error as a function of training data size for random forest, Kriging and simple linear regresson on (a) ozone and (b) PM$_{2.5}$ monitoring data.} 	
  \label{fig:CaseStudySize}	
\end{figure}

\subsection{Case Study Results \& Discussion}
\begin{figure}	
  \centering
    \begin{tabular}{c}
      \includegraphics[width=5in]{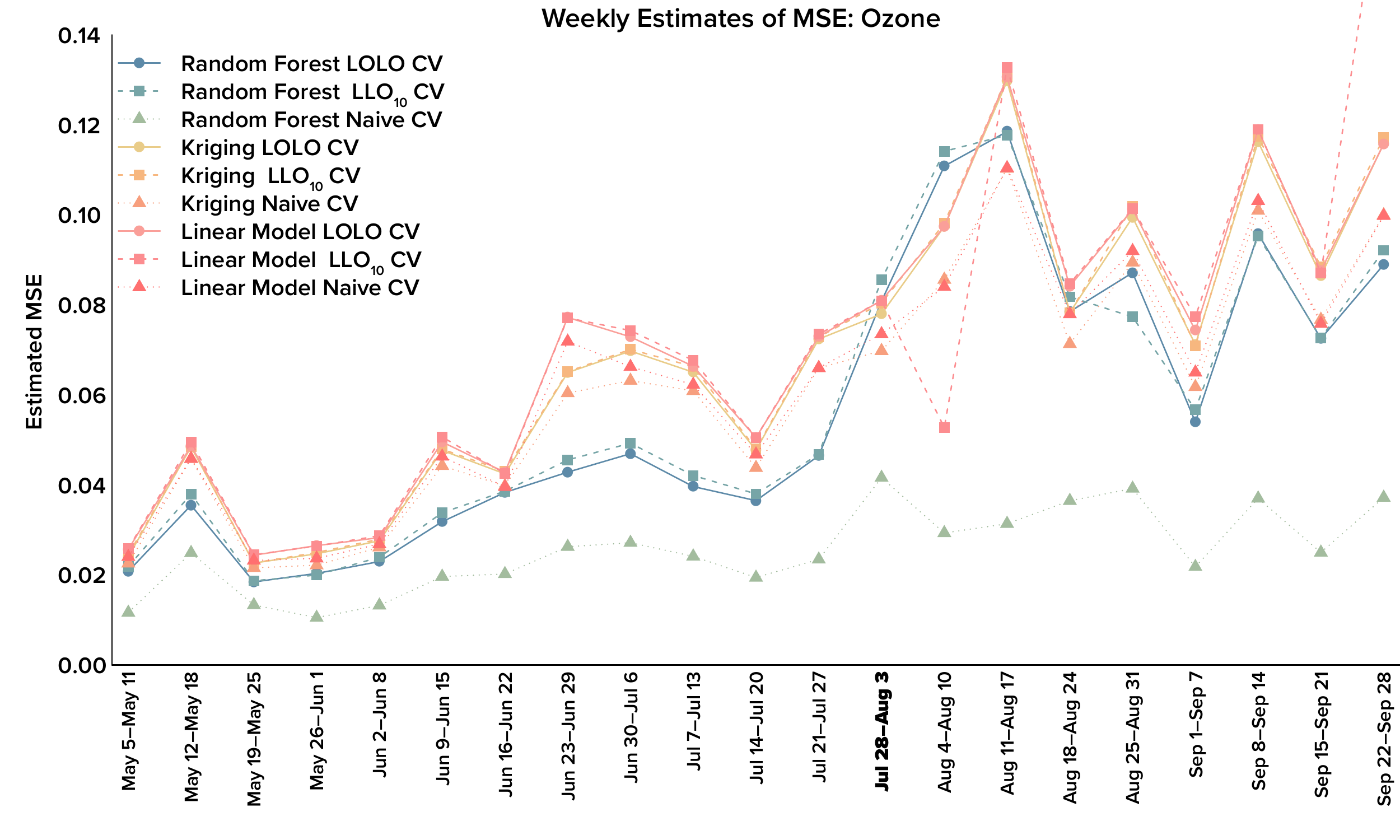}\\
	  (a) Ozone  \\
      \includegraphics[width=5in]{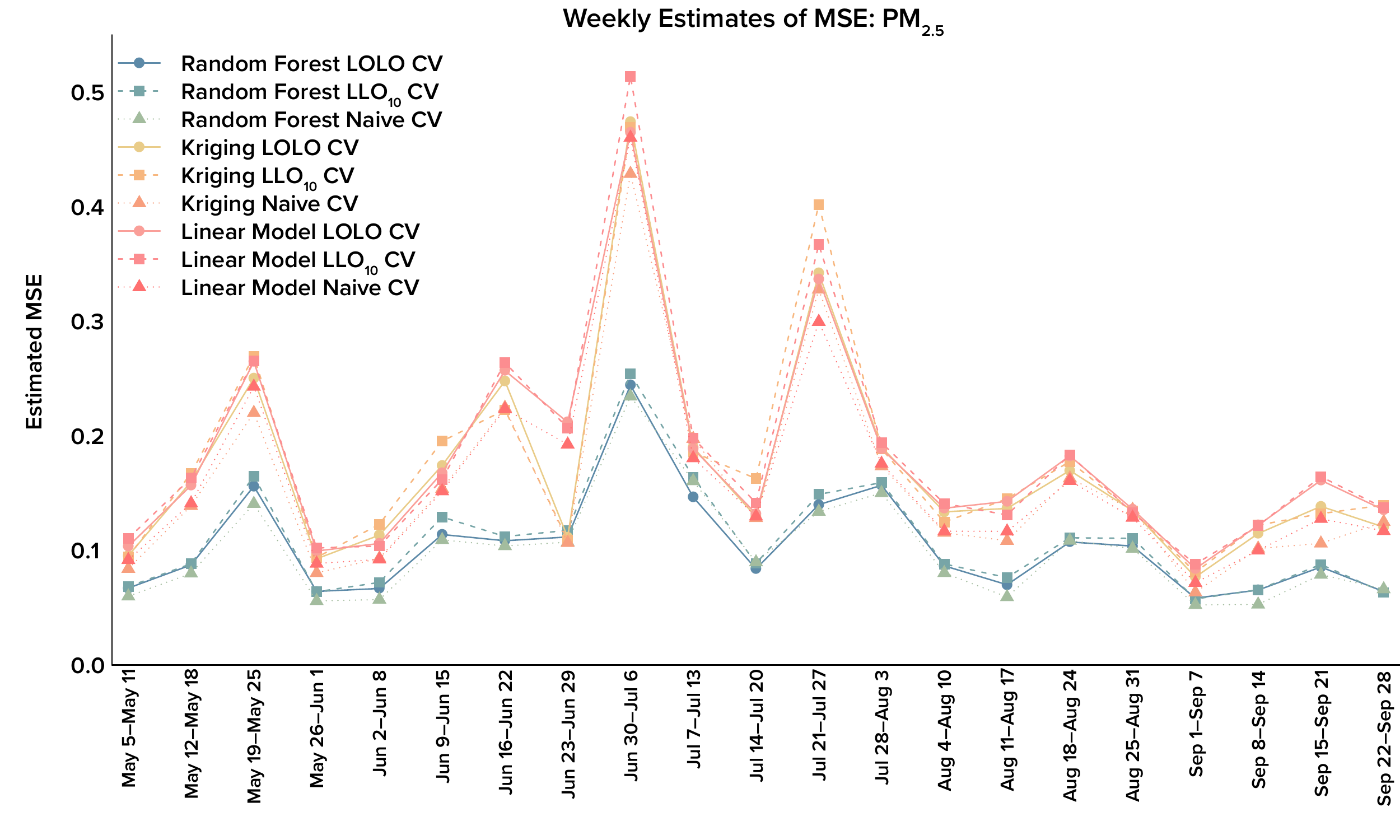}\\
	  (b) PM$_{2.5}$
    \end{tabular}

    \caption{Weekly estimates of prediction error for ozone~(a) and PM$_{2.5}$~(b) monitoring data.} 	
  \label{fig:CaseStudyWeekly}	
\end{figure}

Figure~\ref{fig:CaseStudySize} depicts LOLO, LLO$_{10}$ and naive CV prediction error estimates on the ozone~(a) and PM$_{2.5}$~(b) monitor data for increasingly long subsets of the data. Just as in the simulations, the LOLO and LLO$_{10}$ estimates are similar, and LLO$_{10}$ tends to be slightly more pessimistic. Estimated prediction error generally increases slightly with time series length. While additional timepoints add information, they also increase the prediction burden. This is particularly true if the temporal process is nonstationary, which figure~\ref{fig:CaseStudyWeekly} suggests may be the case at least for the ozone data. 

The naive 10-fold~CV estimates are consistently optimistic compared to the location-based CV estimates. This optimism is considerably more pronounced for random forest than for Kriging or simple linear regression on the ozone data. The naive 10-fold CV estimates are apparently less biased for simple linear regression, because it lacks the flexibility to fully exploit the more highly dependent folds. Kriging, on the other hand, is very flexible, but its performance depends upon accurately estimating the covariance function, which may be quite difficult especially if the process is nonstationary. The optimism of the naive 10-fold~CV ozone estimates for random forest increases with time series length, suggesting that the additional training data at the location where prediction is being made exacerbates the bias. 

Figure~\ref{fig:CaseStudyWeekly} depicts prediction error estimates for models trained and evaluated separately on each weeklong subset of the data. Most of the estimates for ozone increase dramatically in the second half of the study period. This suggests that the process is nonstationary and that the WRF-Chem model may be miscalibrated for this time period, which may both be due to the wildfires that occurred during this time. Interestingly, the naive CV estimates for random forest do not exhibit this trend with estimates staying in approximately the same range across the entire study, clearly indicating once more the dangers of using naive CV for flexible models fit to dependent data. 

Figure~\ref{fig:CaseStudyWeekly} (a) also demonstrates the perils of using naive CV for model selection, averaging or tuning (including variable selection). Consider the week of July 28, 2008, which is bolded along horizontal axis of the figure. Comparing models via naive CV, random forest is the best model by a wide margin with an estimated MSE of 0.042, compared to 0.070 and 0.074 for Kriging and the simple linear regression respectively. In contrast, using LOLO or LLO$_{10}$ Kriging has the lowest MSE. Model ranking is thus not invariant to choice of evaluation metric, which means that naive CV is inappropriate for any procedure that depends upon the relative ranking of predictive performance. These includes model and variable selection, model parameter tuning and model averaging, which often uses relative  performance to compute model weights~\citep{van2007super}.

\section{Discussion}
\label{sec:discussion}

The simulation and case study results confirm that location-based CV techniques are fairly robust estimators of interpolation error over several classes of models, consistent with recent best practice. Naive CV is shown to be highly unreliable for interpolation error estimation as well as for model comparison on space-time data. Buffered CV, which excludes observations within a buffer distance from the training data, is overly pessimistic and appears to target the replication error---the expected error for predicting a new instance of the entire space-time process---rather than the interpolation error, which explicitly conditions upon the observed portions of the space-time field. It also requires selecting the buffer distance, which may be difficult and prompt ad hoc decisions regarding the search window. In addition, even for independent data, CV procedures that include a tuning parameter (e.g., the buffer distance) may not be consistent~\citep{dudoit2005asymptotics}.

LOLO and LLO$_{10}$ CV both appear to target the true interpolation error with LLO$_{10}$ generally displaying a greater pessimistic bias. This is partially due to the larger amount of data in the LOLO training folds, which is analogous to the well-known bias-variance tradeoff in CV fold size for independent data. Reserving more locations from model training as the test fold in LLO$_{10}$ CV eliminates portions of the space-time field from the training data. How this impacts estimates of prediction error depends upon the strength and distribution of the space-time dependence, time series length and spatial sampling intensity. It can be the case that LOLO CV has lower bias \emph{and} lower variance than LLO$_{10}$. Consequently we recommend LOLO CV for estimating interpolation error unless it is computationally infeasible, in which case LLO$_{10}$ is a reasonable alternative. 

This judgment is an empirical one, as there is a lack of theoretical characterization of these procedures in an asymptotic setting. Asymptotics are challenging for CV even in the case of independent data, which admits to replication. In a space-time setting asymptotics may be characterized in multiple ways increasing the difficulty of the problem. 

Random forest, Kriging and simple linear regression were selected because they are popular examples of 3 model classes frequently used for spatial or space-time interpolation. This gives us confidence that the behavior of the estimators observed here will hold across spatially interpolating models more generally. The models may not be the optimal models for these or other interpolation problems. Gradient boosting \citep{watson2019machine, reid2015spatiotemporal} and sometimes deep learning neural networks have outperformed random forest in previous studies. Kriging models with more flexible mean structures \citep{berrocal2010spatio} or less stringent assumptions on the form of the covariance function may outperform the Kriging models considered here. The ability to reliably compare these and any other interpolation techniques is precisely the motivation behind this work, and the consistent relationship between CV estimators of prediction error demonstrated here will likely hold across a wide variety of models. Crucially this includes methods that accommodate high dimensional data. This is particularly important for Kriging models, which have traditionally been limited to relatively small data sets due to the matrix inversion involved, which does not scale well either in computations ($O(n^3)$) or memory usage ($O(n^2)$)~\citep{cressie2015statistics}. There are now a variety of computationally expedient approaches that have improved the scaleability of Kriging-style models, including Gaussian process approximations \citep{katzfuss2017multi} and nearest neighbor Gaussian process models \citep{datta2016hierarchical}. See \cite{heaton2019case} for a recent summary and comparison of these and other approaches. 

More useful than the specific guidance this work suggests for the appropriate choice of CV techniques for evaluating interpolation models is its formal discussion of prediction error estimation for space-time data. While there has been growing acknowledgment in the literature that naively applying unstructured CV on space-time data is ill advised, there is a lack of clarity on what prediction error means in the presence of space-time dependence. Nowhere is this more obvious than in the recent trend towards reporting naive, location-based and timepoint-based CV estimates of prediction error while offering no guidance or discussion on which is appropriate for the problem at hand~\citep{brokamp2018predicting,hu2017estimating}.  

This work fills this void,  distinguishing estimators that target interpolation error from those that appear to target replication, imputation or extrapolation errors.  The appropriate space-time prediction error depends entirely upon the predictive problem at hand. Identifying the relevant space-time prediction error for a particular problem and then selecting an evaluation method for that specific prediction error is the appropriate model evaluation procedure. 

We have not dealt with issues related to consistency of model selection, focusing strictly on prediction error estimation. A model selection procedure is generally described as consistent if the probability it selects the true model goes to 1 as $n \to \infty$. Leave-one-out (LOO) CV is known to be an inconsistent criterion for model selection on independent data, although it can be adapted to reclaim consistency for at least certain classes of models~\citep{shao1993linear}. We do not expect location-based CV procedures to provide consistent selection of the true model. Nevertheless they may be quite useful for selecting/averaging models on the basis of a principled assessment of predictive performance. 
\bibliography{ozone}

\pagebreak
\begin{center}
\textbf{\large {Prediction \& Model Evaluation for Space-Time Data: Supplemental Material}}
\end{center}
\setcounter{table}{0}
\setcounter{section}{0}
\setcounter{page}{1}
\renewcommand*{\thesection}{S\arabic{section}}

\section{Case Study Covariates}~\\

  \small
  \label{tbl:Covariates}
  \begin{tabular}{ll}
    \hline
    Covariate  & Data Source \\
    \hline
Monitor Latitude	& U. S. Environmental Protection Agency \\
Monitor Longitude	& U. S. Environmental Protection Agency \\
Elevation (m)	& National Digital Elevation Model \\
Date	& U. S. Environmental Protection Agency \\
Dew Point ($^{\circ}$K)	& Rapid Update Cycle \\
Boundary Layer Height (m)	& Rapid Update Cycle \\
Surface Pressure (Pa)	& Rapid Update Cycle \\
Relative Humidity (\%)	& Rapid Update Cycle \\
Temperature at 2 m ($^{\circ}$K)	& Rapid Update Cycle \\
U-Component of Wind Speed (m/s)	& Rapid Update Cycle \\
V-Component of Wind Speed (m/s)	& Rapid Update Cycle \\
Inverse Distance to Nearest Fire (m$^{-1}$)	& Fire Inventory from NCAR v1.5 \\
Annual Average Traffic within 1 km	& Dynamap 2000, TeleAtlas \\
Agricultural Land Use within 1 km (\%)	& 2006 National Land Cover Database \\
Urban Land Use within 1 km (\%)	& 2006 National Land Cover Database \\
Vegetation Land Use within 1 km (\%)	& 2006 National Land Cover Database \\
Normalized Difference Vegetation Index	& Landsat Data \\
Nitrogen Dioxide (log molecules/cm$^2$)	& Ozone Monitoring Instrument Satellite \\
WRF-Chem Carbon Monoxide  (log moles/day) & WRF-Chem \\
WRF-Chem PM$_{2.5}$  (log kg/day) & WRF-Chem \\
WRF-Chem Ozone (log 8 Hour Maximum)	& WRF-Chem  \\
    \hline
  \end{tabular}

\section{Simulation Procedure}
Covariates for the simulations described in section 3 were generated from a 6-dimensional Gaussian process (GP) with mean function $\mu_X(s,t)$ and covariance function $\Sigma(s,t)$, i.e., 
\begin{equation*}
  X_1, ..., X_6 \sim GP(\mu_X(s,t), \Sigma(s,t)),
\end{equation*}
where $s$ indexes spatial locations and $t$ indexes time. A combination of sinusoidal functions was selected for the mean function to induce some periodicity across space and time, 
\begin{equation*}
  \mu_X(s,t) = \sin(t) \frac{\sin(s_1)}{250u_1} \frac{\cos(s_2)}{125 u_2} + \cos(t) \left( \frac{\cos(s_1)}{125 u_3} + \frac{\sin(s_2) }{250 u_4} \right), 
\end{equation*}
where $u_1, ..., u_4$ are uniformly distributed random variables, i.e., $u_1, ..., u_4 \sim U[0,1]$. 

The space-time covariance function $\Sigma(s,t)$ was constructed as the product of a spacial covariance function $\Sigma_s$ and a temporal covariance function $\Sigma_t$, i.e., $\Sigma(s,t) = \Sigma_s \Sigma_t$. These were parameterized as squared exponential covariance functions, 
\begin{align*}
  \Sigma_s(s_1, s_2) &= \exp \left(-\frac{\lVert s_1 - s_2 \rVert^2}{v_s^2}\right) + \epsilon I_{n_s}, \\
  \Sigma_t(t_1, t_2) &= \exp \left(-\frac{\lVert t_1 - t_2 \rVert ^2}{v_t^2} \right) + \epsilon I_{n_t},
\end{align*}
where $\lVert \cdot \rVert$ is the Euclidean norm, $v_s$ and $v_t$ scale dependence as a function of distance, and $\epsilon I_{n_s}$ and $\epsilon I_{n_t}$ are diagonal matrices that add a small constant for numerical stability. 

The simulation outcome $y$ was simulated from a GP with mean function $\mu_y(s,t)$ and covariance function $\Sigma(s,t)$,
\begin{equation}
  y \sim GP(\mu_y(s,t), \Sigma(s,t)),
\end{equation}
where 
\begin{equation*}
  \mu_y(s,t) = 0.5 x_1 - 0.5 I(x_2 > 0.1) + (1 + \exp(-x_3))^{-1}.
\end{equation*}
The outcome is independent of the remaining 3 covariates, $x_4$, $x_5$, and $x_6$, which were included as spurious covariates to assess the robustness of prediction models.

\section{Simulation Parameters}~\\

\begin{tabular}{c c c c c c c}
  Scenario & Replicates    & \makecell{Observed\\Locations} & \makecell{Spatial \\ Dependence}  & \makecell{Temporal\\Dependence} & Locations & Days \\
  \hline
1 &  100 &  50  & Low      & Moderate &  400 &  10 \\ 
2 &  100 &  50  & Low      & Moderate &  400 &  20 \\
3 &  100 &  50  & Moderate & Moderate &  400 &  10 \\
4 &  100 &  50  & Moderate & Moderate &  400 &  20 \\
5 &  100 & 150  & Low      & Moderate &  400 &  10 \\
6 &  100 & 150  & Low      & Moderate &  400 &  20 \\
7 &  100 & 150  & Moderate & Moderate &  400 &  10 \\
8 &  100 & 150  & Moderate & Moderate &  400 &  20 \\ \hline
\end{tabular}
\label{SimScenarios}

\section{Conditional Variance}

Simulation outcome $y(s,t)$ was simulated over a space-time lattice from a Gaussian Process with mean function $\mu_y(s,t)$ and covariance function $\Sigma(s,t)$ (see Section~S2). Let $\mathbf{y} = \{y(s_1, t_1), \ldots, y(s_n, t_m)\}$ denote $y$ over the $n$ spatial locations and $m$ timepoints of the simulation lattice, and $\Sigma_{\mathbf{y}}$ its covariance matrix.
Let $\mathbf{y}^*$ denote any subset of $\mathbf{y}$ and $\mathbf{y}^c$ its complement, i.e., $\mathbf{y}^c = \mathbf{y}\setminus \mathbf{y}^*$. 
The covariance matrices of $\mathbf{y}^*$ and $\mathbf{y}^c$ are $\Sigma_{\mathbf{y}^*}$ and $\Sigma_{\mathbf{y}^c}$ respectively, which are submatrices of $\Sigma_{\mathbf{y}}$ with the rows and columns corresponding to $\mathbf{y}^*$ and $\mathbf{y}^c$ respectively.
The conditional covariance matrix of $\mathbf{y}^*$ given $\mathbf{y}^c$ is
\begin{equation*}
    \Sigma_{\mathbf{y}^* \mid \mathbf{y}^c} = \Sigma_{\mathbf{y}^*} - \Sigma_{\mathbf{y}^* \mathbf{y}^c} \Sigma_{\mathbf{y}^c}^{-1} \Sigma_{\mathbf{y}^*\mathbf{y}^c }', 
\end{equation*}
where $\Sigma_{\mathbf{y}^* \mathbf{y}^c}$ is the submatrix of $\Sigma_{\mathbf{y}}$ with rows corresponding to $\mathbf{y}^*$ and columns to $\mathbf{y}^c$. The conditional variance of $\mathbf{y}^* \mid \mathbf{y}^c$ is the diagonal of $\Sigma_{\mathbf{y}^* \mid \mathbf{y}^c}$.

\end{document}